\documentclass[amsmath,amssymb,12pt]{revtex4}
\usepackage{natbib}

\usepackage{epsfig}
\usepackage{graphicx}
\usepackage{amsmath}
\usepackage{amssymb}

\newcommand{\Sch}{{Schr\"{o}dinger}}

\newcommand{\W}   {{\Omega}}
\newcommand{\w}   {{\omega}}
\newcommand{\G}   {{\Gamma}}
\renewcommand{\c} {{\cos\theta}}
\newcommand{\s}   {{\sin\theta}}

\newcommand{\1}       {{|1\rangle}}
\newcommand{\2}       {{|2\rangle}}

\newcommand{\D} {{\Delta}}

\newcommand{\E}   {{\cal E}}

\begin{document}

\title{Pulsed Adiabatic Photoassociation via Scattering Resonances}

\author{Alex C. Han$^1$, Evgeny A. Shapiro$^2$,
 Moshe Shapiro,$^{1,2}$
~\\\it Departments of Physics$^1$ and Chemistry$^2$, The University of British
Columbia
\\  2036 Main Mall, Vancouver, BC, Canada V6T 1Z1}

\begin{abstract}
We develop the theory for the Adiabatic Raman Photoassociation
(ARPA) of ultracold atoms to form ultracold molecules in the
presence of scattering resonances. Based on a computational method
in which we replace the continuum with a discrete set of
``effective modes'', we show that the existence of resonances
greatly aids in the formation of deeply bound molecular states. We
illustrate our general theory by computationally studying the
formation of $^{85}$Rb$_2$ molecules from pairs of colliding
ultracold $^{85}$Rb atoms. The single-event transfer yield is
shown to have a near-unity value for wide resonances, while the
ensemble-averaged transfer yield is shown to be higher for narrow
resonances. The ARPA yields are compared with that of (the
experimentally measured) ``Feshbach molecule''
magneto-association. Our findings suggest that an experimental
investigation of ARPA at sub-$\mu$K temperatures is warranted.


\end{abstract}

\maketitle

\section{Introduction}
\begin{figure}[top]
\centering
\includegraphics[scale=0.45]{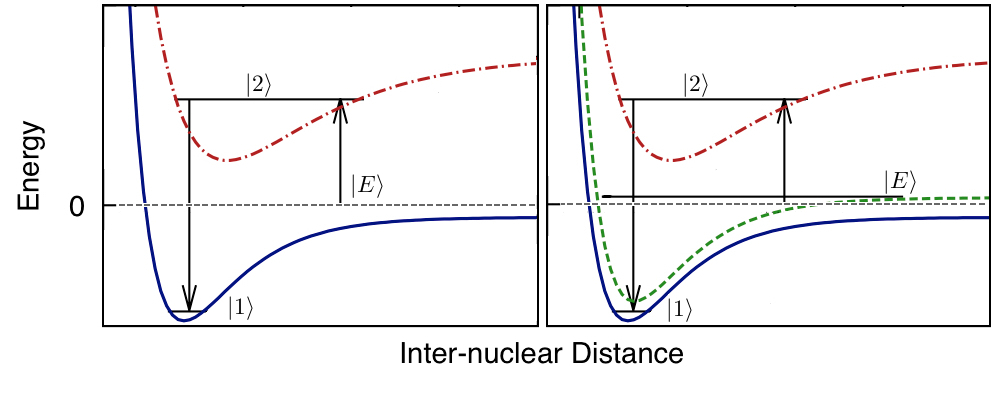}
\caption{A schematic display of the Adiabatic Raman
Photoassociation (ARPA) process. {\bf Left panel:} Atoms colliding in
the near-threshold energy range are excited by the ``pump'' laser
to the vibrational state $|2\rangle$ on an excited electronic
potential. The latter is coupled by the ``dump'' laser to the
deeply bound target state $|1\rangle$.
{\bf Right panel:} The same as on the left panel, with the
continuum-bound couplings modified due to presence of a Feshbach
resonance.} \label{potentials}
\end{figure}

Adiabatic Raman Photoassociation (ARPA) of ultracold atoms was
introduced \cite{vardi,coldbook,zhenia07} as a practical way of
producing ultracold diatomic molecules in their ground
electronic and vib-rotational
states. As illustrated in Fig.\ref{potentials},
the method consists of photoassociating two colliding atoms by
two (``dump'' and ``pump'')
laser pulses that are mutually coherent and partially overlapping
in space and time. As in three-level
``Stimulated Raman Adiabatic Passage''
(STIRAP) \cite{stirap1, stirap2, stirap3, stirap4, stirap5},
one uses the ``counter-intuitive''
pulse ordering \cite{stirap1} in which the ``dump'' pulse, connecting the final bound state to an intermediate excited bound state,
precedes the ``pump'' pulse, connecting the continuum to the latter state.
In this way one executes a smooth ``Adiabatic Passage''
from an (ultracold atom-scattering) continuum to
deeply bound molecular states \cite{vardi,coldbook,zhenia07}.

The main problem with the above approach is the
small ``Franck-Condon'' overlap factors between the intermediate bound state
and the continuum. The introduction of a (``Feshbach'') resonance
which is expected to better overlap with the intermediate bound state
can alleviate this problem \cite{Cote-08,elena}.
As we show below, a Feshbach resonance also induces
an important {\it dynamic effect} of prolonging the
lifetime of the spatial region (the ``Franck-Condon window'')
in which photoassociation occurs. In this way a larger fraction of
the atom pairs becomes available for photoassociation.
This aspect has not been considered in previous works on photoassociation via
Feshbach resonances \cite{elena}, where it was concluded
that wide resonances are more effective in promoting photoassociation
than narrow ones.
While we agree with Ref. \cite{elena} concerning the outcome of a single pair
collision, we differ in our conclusions regarding the ensemble
averaged yields:
Based on our calculation of the number of recombining atoms per
laser pulse-pair, we find that narrow resonances are in fact more
effective than wide ones, because by prolonging the lifetime of the Franck
Condon window, narrow resonances allow for more recombination events to occur.
This mechanism more than compensates for the smaller energetic
widths of the narrow resonances.

The structure of this paper is as follows: in section II we develop
the working equations for the ARPA process, based
on representing the continuum as a discrete set of
``effective modes'' \cite{Moshe-reversibility-PRA96}.
In section III we present calculations of
the ARPA population dynamics for a single pair of ultracold $^{85}$Rb atoms,
and demonstrate how the FC window lifetime and spatial extent are being extended
by the resonance.
In section IV we compare the yield of ARPA with that of
magneto-association
\cite{magneto1, magneto2, magneto3, magneto4, magneto5}
in which a pair of atoms in a Feshbach resonance are transformed into a
stable molecule by sweeping over an external magnetic field, thereby
pushing the resonance energy to lie below
the molecular dissociation limit. We show that the two schemes
lead to similar scaling of molecular production yield, but that
ARPA is expected to be more efficient.
Finally in the Appendix we show that the action of the pulses is tantamount to
a quantum projective measurements on the initial continuum wave packet \cite{coldbook,zhenia07}. Concluding remarks are provided in section V.


\section{Theory}
\subsection{The basic formulation}
As illustrated in Fig.\ref{potentials},
ARPA involves a $\Lambda$-type level structure, similar to 3-state STIRAP,
in which two, mutually coherent,
temporally and spatially overlapping, laser
pulses induce adiabatic passage from a molecular
continuum (representing two colliding atoms)
to the target bound level $\1,$ using
an excited bound state $\2$ as an intermediate.
The Hamiltonian of the system is written as (in atomic
units),
\begin{eqnarray}
\hat{H} &=& \hat{H_0}-2\hat{\mu}\sum_{n=1,2} \epsilon_{n}(t)\cos\omega_n t \label{Hfull}\\
{\rm where}~~~~\hat{H_0} &=& E_1|1\rangle\langle 1|+E_2|2\rangle\langle 2| +
\int_{E_{th}}^{\infty}E|E\rangle\langle E|dE,
\end{eqnarray}
is the ``material'' Hamiltonian. The bandwidth of the pulses ranges from being of order of 100$\mu$K, down to tens of nK, which, when compared to the vibrational energy separation, makes it valid to include no other bound states than $|1\rangle$ and $|2\rangle$. The second term in Eq.(\ref{Hfull}) describes the interaction
of $\hat{\mu},$ the transition dipole moment, with the
``dump'' ($n=1$) and ``pump'' ($n=2$) laser pulses, whose respective amplitudes
and central frequencies are $\epsilon_n(t)$
and $\omega_n$. We tune $\omega_2$, the pump center frequency, to be
in near resonance with $\omega_{2,E},$ the transition frequency between
the intermediate state $|2\rangle$ and the continuum states $|E\rangle$; and
$\omega_1,$ the dump center frequency, to be in resonance with
$\omega_{2,1},$ the intermediate-to-final-state transition frequency.
We assume that the laser fields do not vary significantly
over the range of atom-atom distances in which photoassociation occurs,
thereby justifying the elimination of the spatial variation of the fields.

As state $|1\rangle$ we choose a deeply bound rovibrational level belonging to
the ground electronic potential. Given this choice, the
intermediate state $|2\rangle,$
taken to belong to an excited electronic potential, is chosen
to be a vibrational state that overlaps well with the $|1\rangle$
state. The main feature of the continuum we explore here is the embedding of
a (Feshbach-type) resonance, leading to a sharp
energy dependence of continuum-bound transition-dipole matrix elements
$\mu_{2,E}=\langle 2|\hat{\mu}|E\rangle$ \cite{fano,Cote-08,elena}.

Expanding the time-dependent system wave function in the material basis
set,
\begin{equation}
\label{totalstate}
|\Psi(t)\rangle = \sum_{i=1,2} b_i(t) e^{-iE_i t}|i\rangle + \int_{E_{th}}^{\infty}dEb_E(t)e^{-iEt}|E\rangle,
\end{equation}
we obtain, using the time-dependent \Sch's equation
$i\frac{d}{dt}|\Psi(t)\rangle = \hat{H}|\Psi(t)\rangle,$
the orthonormality of the material states
and the Rotating Wave Approximation (RWA), that,
\begin{eqnarray}
\dot{b_1}(t)&=& i\Omega^*_1(t)b_2(t)\\ \label{b2original}
\dot{b_2}(t)&=& i\Omega_1(t)b_1(t)-\Gamma_f b_2(t)+i \int_{E_{th}}^{\infty}\Omega_E(t)b_E(t)  e^{i\Delta_Et}dE\\ \label{beoriginal}
\dot{b_E}(t)&=& i\Omega_E^*(t)b_2(t) e^{-i\Delta_Et},
\end{eqnarray}
where $E_{th}$ is the continuum threshold energy,
$\Delta_E=E_2-E-\omega_2 $ and $ \Delta_{1} = E_2-E_1-\omega_1 $
are the detunings of the pulses, and
$\Gamma_f$ is the spontaneous decay rate of state $|2\rangle.$
There are two Rabi frequencies in the problem, $ \Omega_1(t) =
\epsilon_1(t)\mu_{2,1}e^{i\Delta_{1} t}$, and $\Omega_E(t) =
\epsilon_2(t)\mu_{2,E}.$

Eq. (\ref{beoriginal}) for $b_E(t),$
representing a continuous set of equations
for the continuously varying $E,$ are numerically difficult to solve.
We therefore eliminate Eq. (\ref{beoriginal})
by integrating $b_E(t)$ in time,
\begin{equation}
b_E(t)= b_E(0)+i\int_0^tdt'\Omega_E^*(t')b_2(t') e^{-i\Delta_Et'},
\end{equation}
and substitute this formal solution into Eq. (\ref{b2original}) to obtain that,
\begin{eqnarray}
\dot{b_2} (t) =&& i\Omega_1(t) b_1(t) -\Gamma_f b_2(t) + i\int_{E_{th}}^{\infty}\Omega_E(t)b_E(0)e^{i\Delta_Et} dE\nonumber\\
&& -\epsilon_2(t)\int_{-\infty}^{\infty}       \bigg[
|\mu_{2,E}|^2\int_0^t \epsilon_2(t')b_2(t') e^{-i\Delta_Et'
}dt'\bigg] e^{i\Delta_Et} dE.
\end{eqnarray}
By defining $f_{source}(t),$ the source function,
and $F(t-t'),$ the spectral auto-correlation function as
\begin{equation}
f_{source}(t) = \int_{E_{th}}^{\infty}\Omega_E(t)b_E(0)e^{i\Delta_Et} dE,
\label{fsource}
\end{equation}
\begin{equation}
F(t-t') = \int_{E_{th}}^{\infty}|\mu_{2,E}|^2
e^{i\Delta_E(t-t')}dE, \label{Fcorr}
\end{equation}
we can transform the above (continuous)
set of differential equations to
a set of two integro-differential equations,
\begin{eqnarray}
\dot{b_1}(t)&=& i\Omega^*_1(t)b_2(t)\nonumber \\
\label{b2integral}
\dot{b_2}(t)&=& i\Omega_1(t) b_1(t) -\Gamma_f b_2(t) + if_{source}(t)- \epsilon_2(t) \int_0^t \epsilon_2(t')b_2(t') F(t-t') dt'.
\end{eqnarray}
The threshold energy $E_{th}$ in the source function will later in
our analysis be taken as $-\infty$, since the function $b_E(0)$
is zero near the collision threshold of the ground electronic
potential, reflecting the density of states at zero kinetic
energy.

\subsection{The effective modes expansion}

The simplest solution of Eqs. (\ref{b2integral}) is
obtained by
the ``flat continuum'' or ``slowly varying continuum approximation'' (SVCA),
according to which, whenever $\mu_{2,E}$ varies sufficiently
slowly with energy $E$ we replace it
by its value at some average energy $\overline{E}.$
In this way the spectral auto-correlation function
of Eq. (\ref{Fcorr})
is reduced to $F(t-t') = 2\pi|\mu_{2,\overline{E}}|^2 \delta (t'-t),$ and
the integration in Eq. (\ref{b2integral}) is eliminated.
Given this approximation,
the dynamical equations assume, in matrix notation, the form,
\begin{equation}
\frac{d}{dt}\mathbf{b} = i \textmd{H} \cdot\mathbf{b} +
i\mathbf{f}_{source} \label{SVCA-bdot}
\end{equation}
where $\mathbf{b}(t)\equiv (b_1(t),b_2(t))^{\rm T}$,
$\mathbf{f}_{source}(t)\equiv (0,f_{source}(t))^{\rm T}$,
with $^{\rm T}$ designating the {\it transpose} operation.
The Hamiltonian matrix is defined as,
\begin{equation}
\textmd{H} = \begin{pmatrix} 0& \Omega_1^*\\ \Omega_1&
i\Gamma_{eff}(t) \end{pmatrix} \quad\text{with}\quad
\Gamma_{eff}(t) =\pi|\mu_{2,\overline{E}}|^2\epsilon_2^2(t).
\label{SVCA-H}
\end{equation}
A detailed discussion of the solutions under SVCA was made in Refs.
\cite{vardi,coldbook,zhenia07}.

The SVCA is however invalid when collisional resonances are embedded
in the continuum, because in that case $\mu_{2,E}$
changes rapidly near the resonance energy \cite{fano,Cote-08}.
In order to treat this case we parametrize $\mu_{2,E}$
as \cite{Moshe-reversibility-PRA96}
\begin{equation}
\mu_{2,E}=\sum_{s=1}^M\frac{i\mu_s\Gamma_s/2}{E-E_s+i\Gamma_s/2}
\label{resonances}
\end{equation}
where $\mu_s$ represents the electronic transition dipole moment;
$\Gamma_s$ - the Full-Width-at-Half-Maximum (FWHM);
and $E_s$ - the centre-position of each $s$ resonance.
This form is capable of approximating well both wide and
narrow resonances \cite{fano, elena,Xuan11}. As will be seen below,
the above parametrization allows us to greatly
simplify both the analytical theory as well as
the numerical propagation of the dynamical equations.

With the expansion (\ref{resonances}), the auto-correlation
function $F(t-t')$ in Eq. (\ref{Fcorr}) can be evaluated analytically as,
\begin{equation}
F(t-t') = \sum_{s=1}^M \alpha_s f^+_s(t) f^-_s(t')
\label{fcores}
\end{equation}
with
\begin{eqnarray}
\alpha_s = \sum_{s'} \frac{-i\mu_s\mu_{s'}\Gamma_s\Gamma_{s'}/4}{E_s-E_{s'}-i(\Gamma_s+\Gamma_{s'})/2},~~ f^{\pm}_s(t) = \sqrt{2\pi}e^{\mp i\chi_s t},~~
\chi_s = E_s-E_2+\omega_2-i\frac{\Gamma_s}{2}.
\end{eqnarray}
Using Eq. (\ref{fcores}) we now define \cite{Moshe-reversibility-PRA96}
the ``effective modes'' variables as,
\begin{equation}B_s^-(t) =i\int_0^t\epsilon_2(t') b_2(t') f^-_s(t')dt',
\label{Bdefinition}
\end{equation}
using which, we transform Eqs. (9) and (10) into,
\begin{eqnarray}
\dot{b_1}(t) &=& i\Omega_1^*(t)b_2(t)  \label{resonance-b1dot}\\
\dot{b_2}(t) &=& i \Omega_1(t)b_1(t)-\Gamma_f b_2(t) + if_{source}(t) +i\epsilon_2(t)\sum_{s=1}^M \alpha_sf^+_s(t) B_s^-(t)  \label{resonance-b2dot} \\
\dot{B}_s^-(t) &=& i \epsilon_2(t)f^-_s(t) b_2(t),~~~s=1,...,M.
\label{resonance-Bdot}
\end{eqnarray}
In this way the original set of continuous
equations for $ b_E(t) $ is replaced by a discrete set of
equations for $B^-_s(t)$.
We can further simplify the structure of the equations by
defining
$B_s(t) = \sqrt{{\alpha_s}/{2\pi}}f^+_s(t) B^-_s(t)$, and
mode-dependent Rabi frequencies,
$\Omega_2^{(s)} \equiv\epsilon_2(t) \sqrt{2\pi\alpha_s} $. With these definitions
Eqs. (\ref{resonance-Bdot}) assume the form,
\begin{eqnarray}
\label{b1effmod}
\dot{b_1}(t) &=& i\Omega_1^*(t)b_2(t)\\
\label{b2effmod}
\dot{b_2}(t) &=& i \Omega_1(t)b_1(t) - \Gamma_f b_2(t)+ if_{source}(t) +\sum_s i\Omega_2^{(s)}(t)B_s(t)\\
\label{bseffmod}
\dot{B}_s(t) &=& -i \chi_sB_s(t) + i \Omega_2^{(s)}(t)b_2(t).
\end{eqnarray}
Writing these equations in matrix notation, we have that,
\begin{equation}
\frac{d}{dt} \mathbf{b} = i\textmd{H}\cdot\mathbf{b}+i\mathbf{f}_{source},
\label{effmode}
\end{equation}
where
\begin{eqnarray}
\mathbf{b}(t)=\begin{pmatrix}  b_1(t) \\  b_2(t) \\ B_1(t) \\ B_2(t) \\ \vdots \end{pmatrix}\quad
\mathbf{f}_{source}(t)= \begin{pmatrix} 0\\ f_{source}(t)\\ 0 \\0 \\ \vdots\end{pmatrix}~~{\rm and}~~
\textmd{H}=\begin{pmatrix}
0 & \Omega_1^* & 0  & 0 & \cdots \\
\Omega_1 & i\Gamma_f & \Omega^{(1)}_2 &  \Omega^{(2)}_2 & \cdots\\
0 & \Omega^{(1)}_2 & -\chi_1 & 0 & \cdots \\
0 & \Omega^{(2)}_2 & 0 & -\chi_2 &  \cdots \\
\vdots & \vdots & \vdots & \vdots &  \ddots \\
\end{pmatrix}.
\end{eqnarray}
The ``effective modes'' amplitudes
$B_s(t)$ thus appear equivalent to some extra bound
states of energies $E_s$ that are coupled by the Rabi frequencies
$\Omega^{(s)}_2(t)$ to state $|2\rangle$, with detuning
$E_s-E_2+\omega_2$ and decay rates $\Gamma_s/2$ as contained in
$\chi_s$.
The non-Hermiticity of the Hamiltonian is due not just
to the decay of the effective modes, appearing as the
imaginary part of $\chi_s$, but also to
the Rabi frequencies $\Omega_2^{(s)}$, which are in general
complex numbers, since the definition of $\alpha_s$ involves a summation
over $s'$, namely
the effective interaction between {\it overlapping} resonances.

\begin{figure}[top]
\centering
\includegraphics[width=1.0\columnwidth]{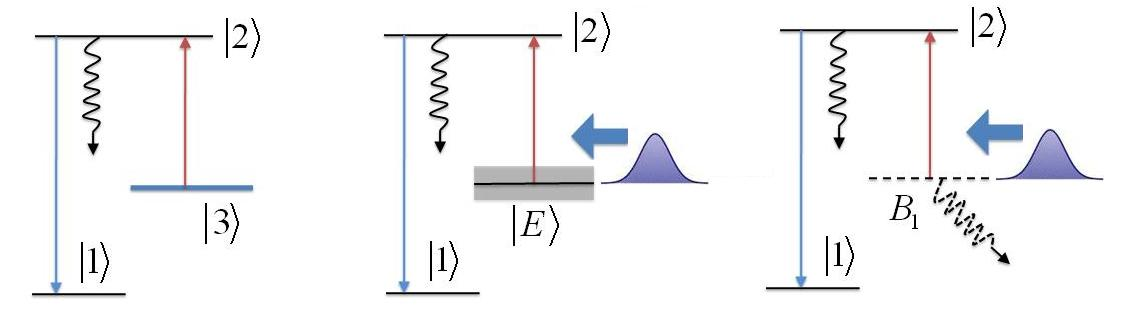}
\caption{{\bf Left panel:} Three bound states STIRAP. The population which
starts in the initial bound state $|3\rangle$ is transferred into state
$|1\rangle$ by following the evolution of the ``dark'' field-dressed state.
The process avoids population loss due to spontaneous emission from
state $|2\rangle$ because the latter remains unpopulated.
{\bf Middle panel:} ARPA via a collisional resonance. The population
gradually feeds the resonances as the continuum wave packet (shaded area)
arrives (at time $t_0$). {\bf Right panel:} The resonance-dominated continuum
of the middle panel is replaced by a single (several) effective mode(s)
with decaying amplitude(s) $B_1(t)$ ($B_s(t)$).} \label{stirap}
\end{figure}

Eq. (\ref{effmode}) resemble (multi-state) STIRAP \cite{ccap}
 with $\Omega_1(t)$ and $\Omega^{(s)}_2(t)$ coupling respectively
$|1\rangle$ with $|2\rangle$, and $|2\rangle$ with each of the
$s$ effective modes
(Fig. \ref{stirap}). We note however that the transfer dynamics
differs in a significant way from ordinary STIRAP in that here the initial
population does not reside in
the effective modes, which get {\it gradually} populated.
We can see this most explicitly for a single resonance, for which
the dynamical equations assume the form,
\begin{eqnarray}
\dot{b_1}(t) &=& i\Omega_1^*(t)b_2(t)\\
\dot{b_2}(t) &=& i \Omega_1(t)b_1(t) - \Gamma_f b_2(t)+  i\Omega_2^{(1)}(t)\bigg[ f_{source}(t)/\Omega_2^{(1)}(t) + B_1(t)\bigg] \\
\dot{B}_1(t) &=& -i \chi_1 B_1(t) + i \Omega_2^{(1)}(t)b_2(t).
\end{eqnarray}
By re-defining $B(t)= f_{source}(t)/\Omega_2^{(1)}(t) + B_1(t)$
we obtain
\begin{eqnarray}
\dot{b_1}(t) &=& i\Omega_1^*(t)b_2(t)\\
\dot{b_2}(t) &=& i \Omega_1(t)b_1(t) - \Gamma_f b_2(t)+  i\Omega_2^{(1)}(t)B(t) \\
\dot{B}(t) &=& -i \chi_1 B(t) + i \Omega_2^{(1)}(t)b_2(t) +
\bigg[ i\chi_1 {f_{source}(t)}/{\Omega_2^{(1)}(t)} + f'_{source}(t)\bigg].
\end{eqnarray}
where
$f'_{source}(t)\equiv\frac{i}{\sqrt{2\pi\alpha_1}}
\int_{-\infty}^{\infty} \Delta_E\mu_{2,E} b_E(0)e^{i\Delta_Et} dE.$
We see that here it is the terms in the square bracket that populates the effective mode (see Fig. \ref{stirap}).

Another contrast with three state STIRAP is the
possibility of leakage of
population from the ``dark'' state.
In three states counter-intuitive pulse ordering adiabatic
passage \cite{stirap1}, the population
resides initially in the adiabatic ``dark'' state, which is
a superposition of the initial and target states only.
In that case, the adiabaticity of the pulses guarantees the completeness of the
transfer from the initial to the target state, leaving the intermediate state
unpopulated at all times. Because in our case the effective mode gets populated
in a gradual fashion, the system wave function
may contain non-negligible contributions from
other (``bright'') adiabatic states.
These ``bright'' states have a small overlap with the intermediate state
$|2\rangle,$ causing population to be lost via spontaneous emission.

\section{Computations of resonant photoassociation}

We view the entire ARPA process as a statistical average over
collisions between individual pairs of colliding atoms.
As explained above, we represent such pairs by a set of
(spatially extended) coherent wave packets
arriving at the Franck Condon region at different times.
The choice of coherent wave packets
(rather than plane waves) as the basis of our our computations is
merely a result of our wish to work with $L^2$ normalizeable states.

In this section we examine the above formulation by performing a set of
computations on the resonantly enhanced photoassociation of ultracold
$^{85}$Rb atoms to form $^{85}$Rb$_2$ in its ground vibrational state.
In keeping with our view of the process we divide the computations into two parts: (A) population transfer for each colliding pair,
and (B) population transfer of the thermal
ensemble of pairs of colliding atoms.


\subsection{The single collision transfer yields}

\begin{figure}[top]
\centering
\includegraphics[width=1\columnwidth]{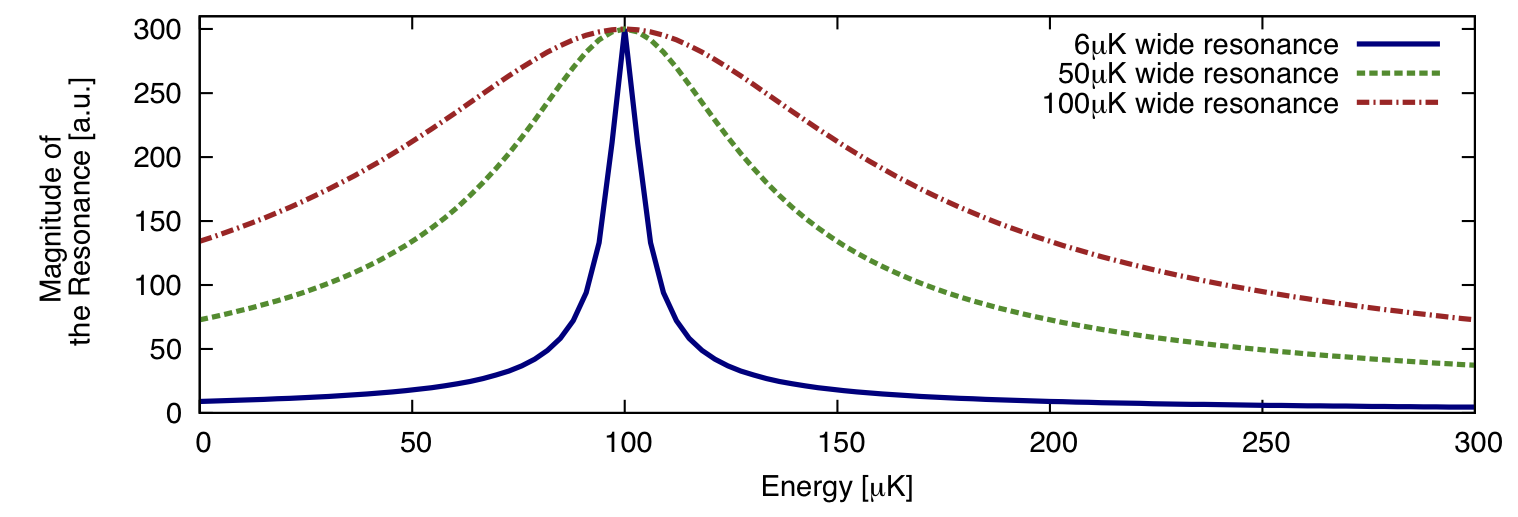}
\caption{A single resonance (in the effective mode expansion) is used for the computation, with magnitude $|\mu_{2,E}| =
\left|\mu_{res}\Gamma_{res}/\left[2(E-E_{res})+i\Gamma_{res}\right]\right|$ as a
function of energy, shown in this figure for various choices of $\Gamma_{res}$. The resonance height has numeric value $\mu_{res}=300$a.u. (depending on the actual physical system, an even larger height value can be used, which favours a lower cost of laser amplitude required). The peaking shape of the resonance represents the enhancement of the FC factor. Centre of the resonance is $E_{res}=100\mu$K.} \label{resonance}
\end{figure}

Following the model of Refs. \cite{coldbook, zhenia07},
we consider a pair of $^{85}$Rb atoms colliding on the ground
electronic potential. We assume that at $t=0$, chosen to occur
before the onset of the pulses,
all the population resides in the continuum wave packet
and none in states $|1\rangle$ or $|2\rangle$.
The shape of
$|\Psi(0)\rangle = \int_{E_{th}}^{\infty}dE b_E(0)e^{-iEt}|E\rangle,$
the initial continuum wave packet of Eq. (\ref{totalstate}),
with $E_{th}$ being the lower energetic limit (which
is extended to $-\infty$),
is determined by the $b_E(0)$ function, chosen here
to be an energetically-narrow
Gaussian \cite{vardi,coldbook,zhenia07}
\begin{equation}
b_E(0) = \frac{1}{(2\pi\delta_0^2)^{1/4}}\exp\bigg[-\frac{(E-E_0)^2}{2\delta_0^2}+i(E-E_0)t_0\bigg],
\label{be0}
\end{equation}
with $\delta_0 = 70 \mu$K and $E_0 = 100\mu$K.
With this choice
of parameters, the wave packet temporal peak occurs at
$t_0 = 1.2 \mu$s \cite{vardi,coldbook}.
The pair of pulses spectral widths are then chosen to have a good overlap with
the energetic spread of the atomic wave packet.
This requirement translates in the time domain to $\mu$s pulse durations.

The scattering continuum is assumed to contain a narrow resonance, whose
shape is given by Eq. (\ref{resonances}).
Figure \ref{resonance} shows the shape of a resonance
centred at $E_{res}=E_0=100\mu$K for 3 different widths.
Since the (Feshbach) resonance position,
can be tuned (magnetically), we can optimize
the transfer by tuning $E_{res}$ to be always equal to $E_0$.
In this way one achieves the maximal FC enhancement (as confirmed
by the numerical calculation presented in Figure \ref{efficiency} bottom
panel).

The bound-to-bound matrix element is chosen to have a numeric value $\mu_{21} = 0.0051$a.u. This numeric value can be different depending on the actual experimental set-up, but our results only depend on the Rabi frequency $\Omega_1(t)$, which is proportional to the product of this bound-to-bound matrix element with the dump pulse amplitude. A increase (decrease) of the matrix element translates into a proportional decrease (increase) in the laser amplitude required. The spontaneous decay rate from level $|2\rangle$ is
$\Gamma_f = (30$ns$)^{-1}$ \cite{vardi,coldbook,zhenia07}.
The central frequency of the
dump pulse is chosen to coincide with $E_2-E_1,$
and the central frequency of the pump pulse - to coincide with $E_2-E_0$.
The field amplitudes
$\epsilon_{1}(t)$ and $\epsilon_{2}(t)$
are taken as Gaussian functions, peaking, respectively,
at $1.05\mu$s and $1.55\mu$s. The duration of both fields is $0.22\mu$s,
and their peak intensity is $3\times 10^5$ W/cm$^2$ \cite{coldbook}.

With these specifications,
Eqs. (\ref{b1effmod}-\ref{bseffmod}) simplify to yield,
\begin{eqnarray}
\label{b1dyneq}
\dot{b_1}(t) &=& i\Omega_1^*(t)b_2(t)\\
\label{b2dyneq}
\dot{b_2}(t) &=& i \Omega_1(t)b_1(t) - \Gamma_f b_2(t)+ if_{source}(t) +i\Omega_2^{(1)}(t)B_1(t)\\
\label{B1dyneq}
\dot{B}_1(t) &=& -(\Gamma_{res}/2) B_1(t) + i \Omega_2^{(1)}(t)b_2(t)
\end{eqnarray}
where the Rabi frequencies are $\Omega_1(t) = \epsilon_1(t)\mu_{2,1}$ and
$\Omega_2^{(1)} = \epsilon_2(t)\mu_{res}\sqrt{2\pi\Gamma_{res}}/2$. Notice here that the FC factors contained in $\mu_{2,1}$ and $\mu_{res}$ always appear as product with the field amplitudes $\epsilon_{1,2}(t)$. So the intensities of the dump and pump fields really are determined by the respective FC factors between the intermediate and target states, and between the continuum and the intermediate state. An enhancement on either of the the FC factors will result in the same order decrease in the laser amplitude (square-root of the intensity) needed.

\begin{figure}[top]
\centering
\includegraphics[width=1\columnwidth]{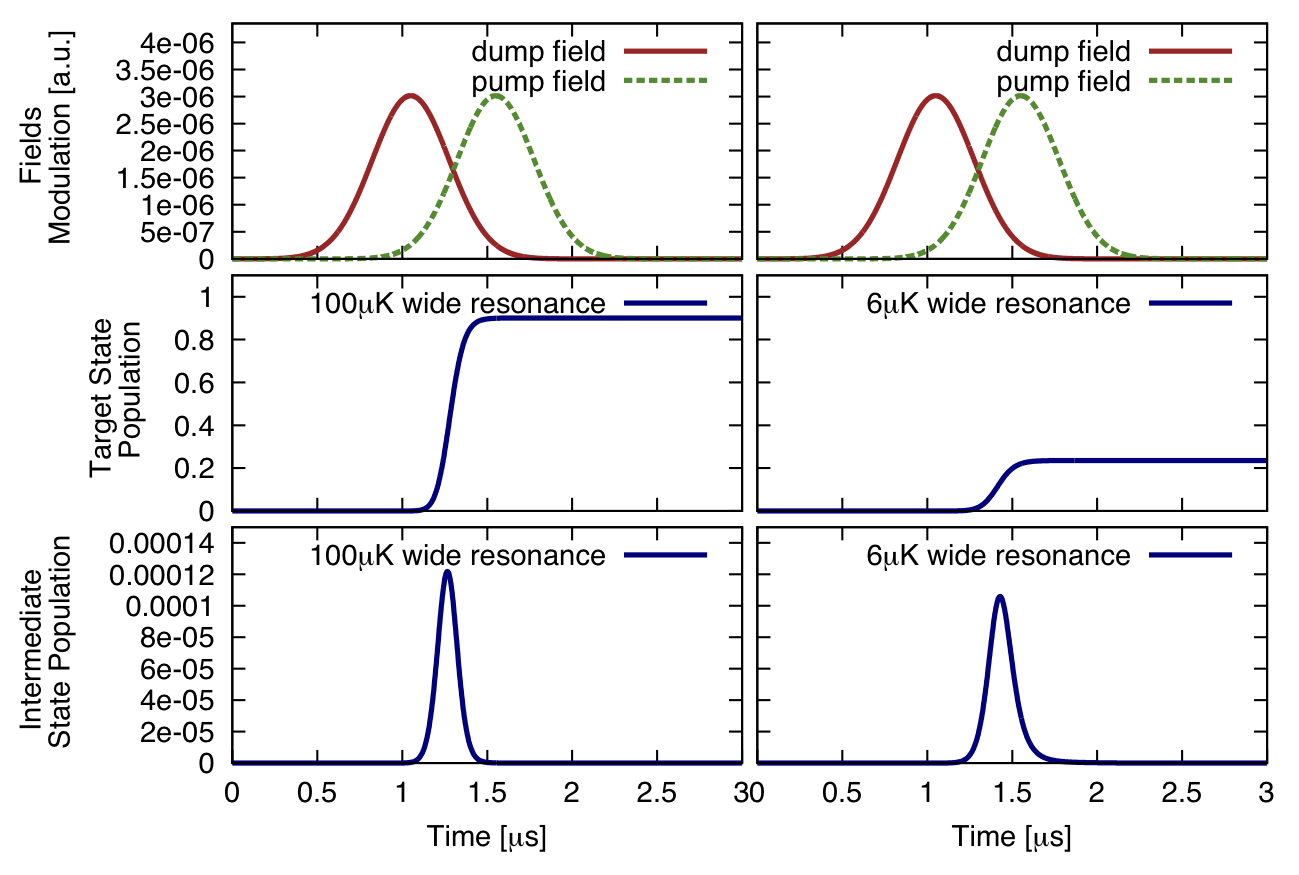}
\caption{The target state and intermediate state
populations as a function of time for
two resonance widths. {\bf Top panel}:
the pump and dump field amplitudes.
{\bf Middle panel}: The target state probability $|b_1(t)|^2$.
The transfer yield is $90\%$ for the wide
($100\mu$K) resonance, but only $23\%$ for the narrow
($6\mu$K) resonance. {\bf Bottom panel}: The intermediate state population
$|b_2(t)|^2$. (Notice the large difference in the vertical scale relative to
the middle panel.)} \label{dynamics}
\end{figure}

In Fig. \ref{dynamics} we display the results
of numerically integrating the equations for $b_1(t)$, $b_2(t)$ and $B_1(t),$
given that $b_1(0) = b_2(0) = B_1(0) = 0$.
We plot the populations of states
$|1\rangle$ and $|2\rangle$ for
a wide ($100\mu$K) resonance and a narrow ($6\mu$K) resonance.
The most striking feature of this plot is that the wide resonance
gives rise to an essentially complete population transfer ($> 90\%$),
while the transfer probability via the narrow resonance is only $\sim 23\%$.
Since the target state is the ground state, no loss of population can occur
after a single photoassociation event. Loss of population is
however possible when subsequent collisions with the gas of atoms
and/or the action of subsequent pulses are considered.
As shown in the lower panel, due to the adiabatic nature of the process and
the ``counter-intuitive'' pulse ordering,
the population of the intermediate
level $|2\rangle,$ remains very low, even while the pulses are on.

\begin{figure}[top]
\centering
\includegraphics[width=1\columnwidth]{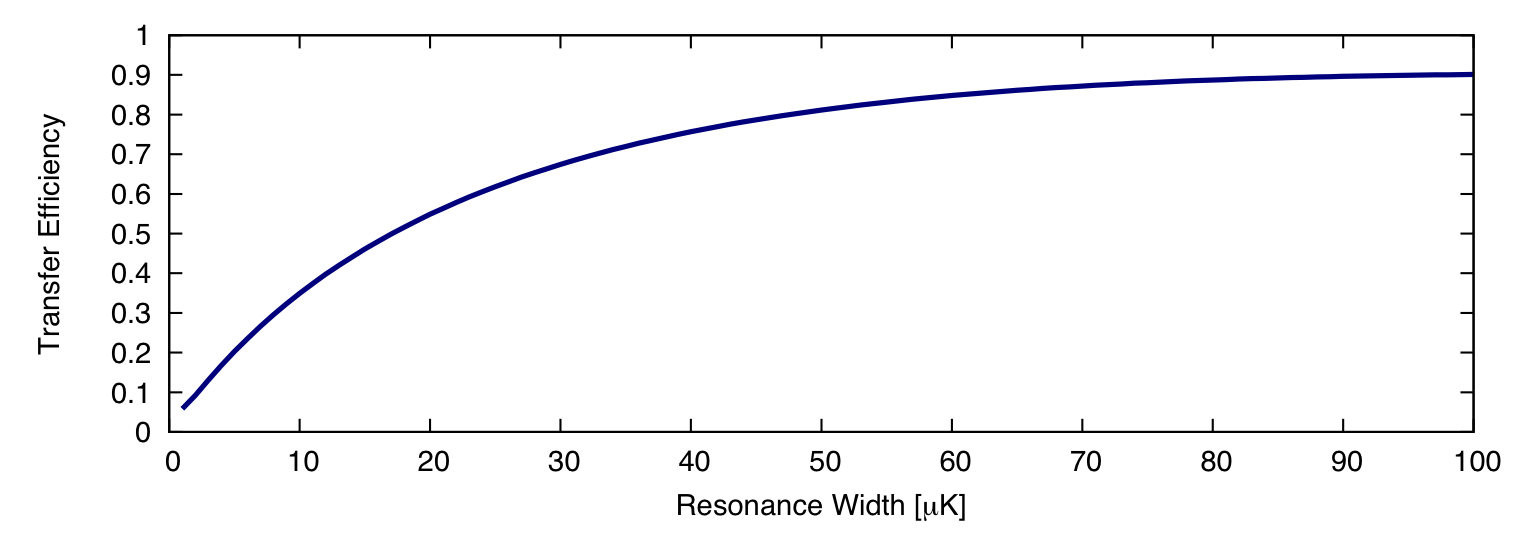}
\includegraphics[width=1\columnwidth]{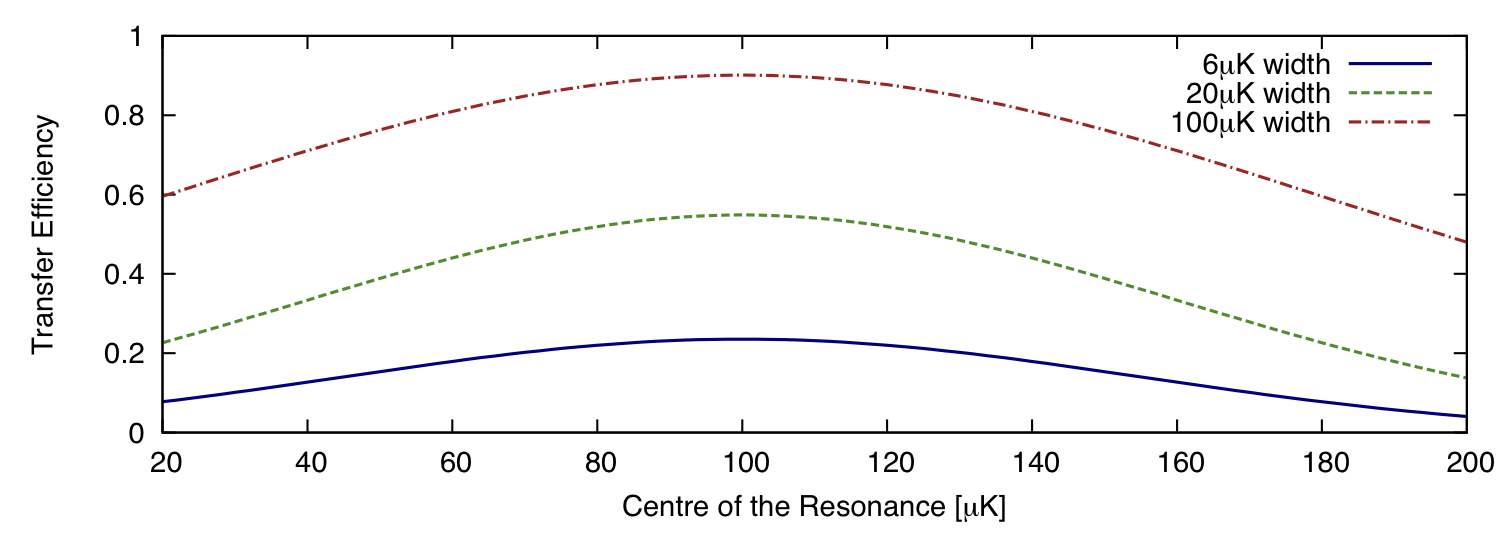}
\caption{{\bf Top panel:} The target population,
$|b_1(t\rightarrow\infty)|^2$, for different resonance widths.
{\bf Bottom panel:} $|b_1(t\rightarrow\infty)|^2$
as a function of the centre of the resonance $E_{res}$, for $E_0=100\mu$K; this shows transfer is optimal when the centre of the resonance coincides with the central energy of $b_E(0)$.}
\label{efficiency}
\end{figure}

\subsection{{Ensemble transfer yield}}

In agreement with Ref. \cite{elena},
we have shown in sub-section A that
for each event the transfer yield via a wide resonances is
greater than that of a narrow resonances (Fig. 5 top panel).
The situation is however different for an {\it ensemble}
of colliding atoms, where, as we show below, the
transfer yield of the {\it narrow} resonances is {\it greater}. The reason is
that for narrow resonances the number of colliding pairs which can react
to the light is greater, essentially because for narrower
resonances one can work with narrower pulse bandwidths, hence longer pulses.
The increase in the number of effective
collisions occurring during the increased pulse durations,
more than compensates for the reduction in the individual event
transfer yield.

An alternative way of viewing this effect is to examine the role of
the discrete effective modes which replace the continuum in our theory.
These modes are to all intents and purposes {\it resonances} \cite{taylor}.
The only difference between the modes and scattering resonances is that
the effective modes do not originate from a {\it real} bound state embedded in a continuum.
Thus, as clearly seen in Eq. (\ref{B1dyneq}), the rate of de-populating an
effective mode is proportional $\Gamma_{res}$ - the resonance-width of that mode.
Hence narrower resonances, corresponding to
smaller rates of depopulation, increase the interaction
times of the effective modes with the intermediate level $|2\rangle$,
thereby prolonging the duration of the Franck-Condon window.

\begin{figure}[top]
\centering
\includegraphics[width=1\columnwidth]{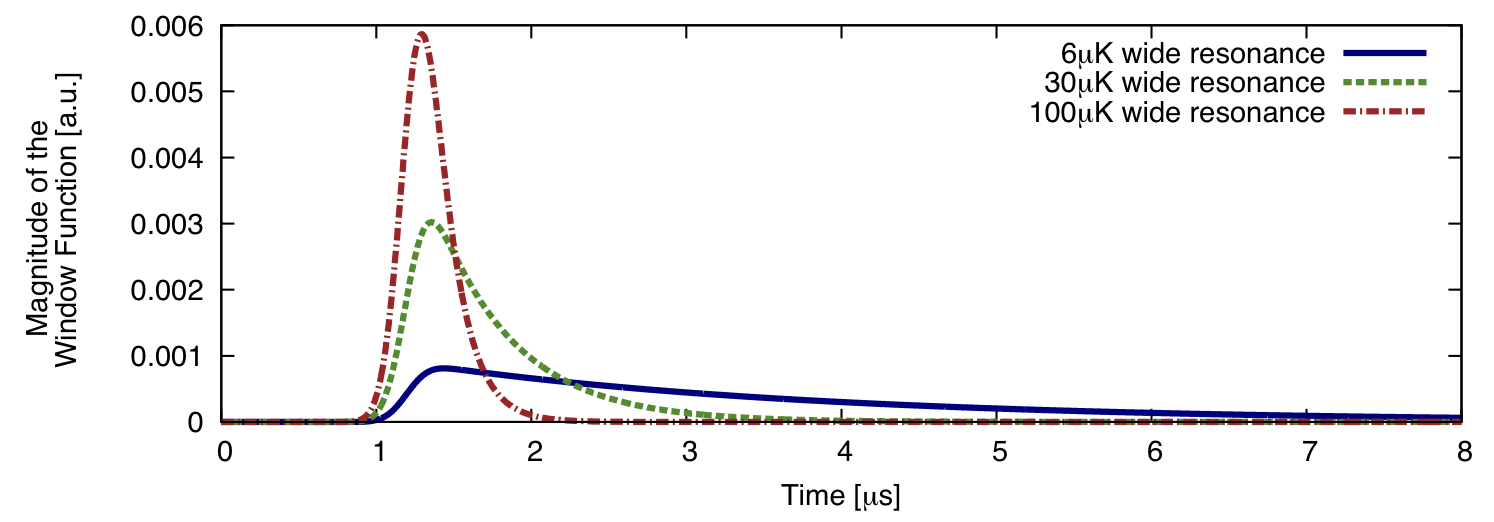}
\caption{The magnitude of the window functions $f^W(t)$ for
different (single) resonance widths $\Gamma_{res}$ and fixed height $\mu_{res}.$
Longer tails of $ |f^W(t)| $ are observed for narrower resonances.}
\label{fwindow}
\end{figure}

In Fig. \ref{fwindow} we examine
these trends in a
quantitative way by displaying $f^W(t),$
the field normalized source term, given as,
\begin{equation}
 f^W(t) = f_{source}(t)/\epsilon_2(t) =
\int_{-\infty}^{\infty} \mu_{2,E}b_E(0)e^{i\Delta_Et} dE,
\end{equation}
for resonances of changing widths.
Clearly in evidence is the prolonged duration of $f^W(t)$ when
switching to narrower resonances.

\begin{figure}[top]
\centering
\includegraphics[width=1\columnwidth]{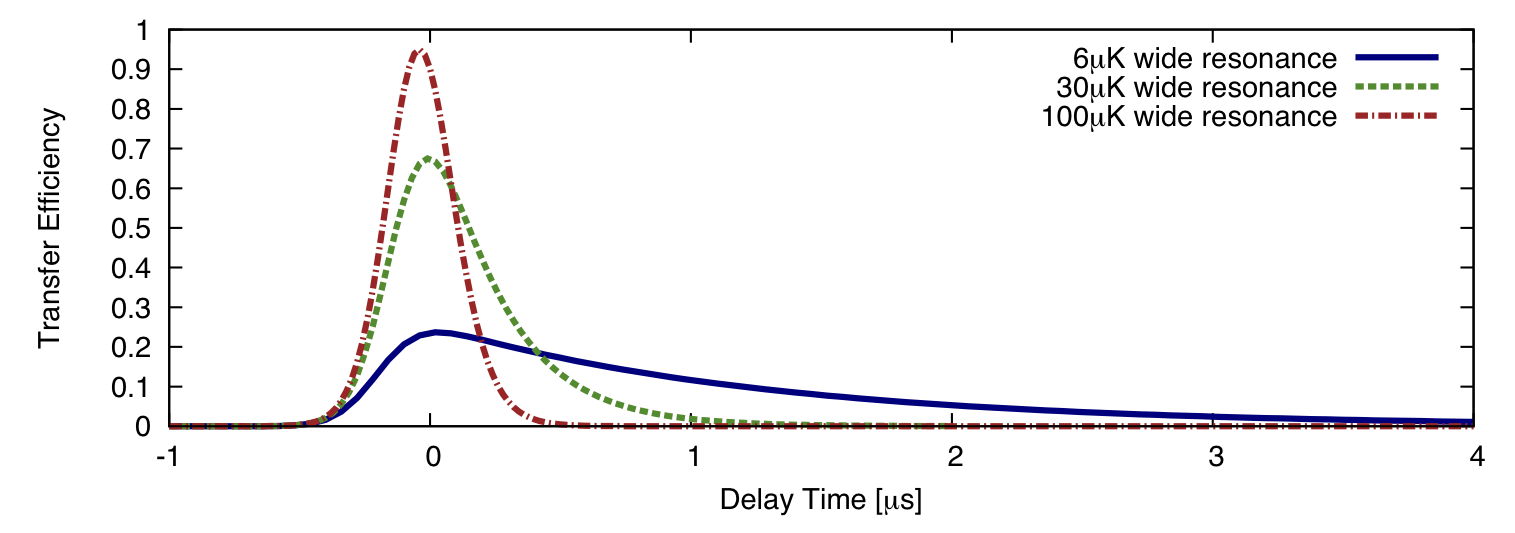}
\caption{The $|b_1(t\rightarrow\infty)|^2$ transfer yield for different
resonance widths as a function of the
\newline $\delta t\equiv t_0-t_P$ delay time,
where $t_P=1.2\mu$s is the pulses' overlap peak time,
and $t_0$ is the incoming wave packet peak time.}
\label{delay}
\end{figure}

The temporally stretched population source is also beneficial
when we consider the action of a pulse pair that is delayed relative to $t_0,$
the arrival time of the incoming wave packet. Figure \ref{delay} shows the
transfer efficiency as a function of such delay times for 3 different
resonance widths. For a narrow resonance, despite the drop in the peak value,
the single collision transfer efficiency remains large for longer times.
This means that atom pairs which started their collision at an earlier time
can still be transformed into bound molecules with non-negligible probability.

\begin{figure}[top]
\centering
\includegraphics[width=1\columnwidth]{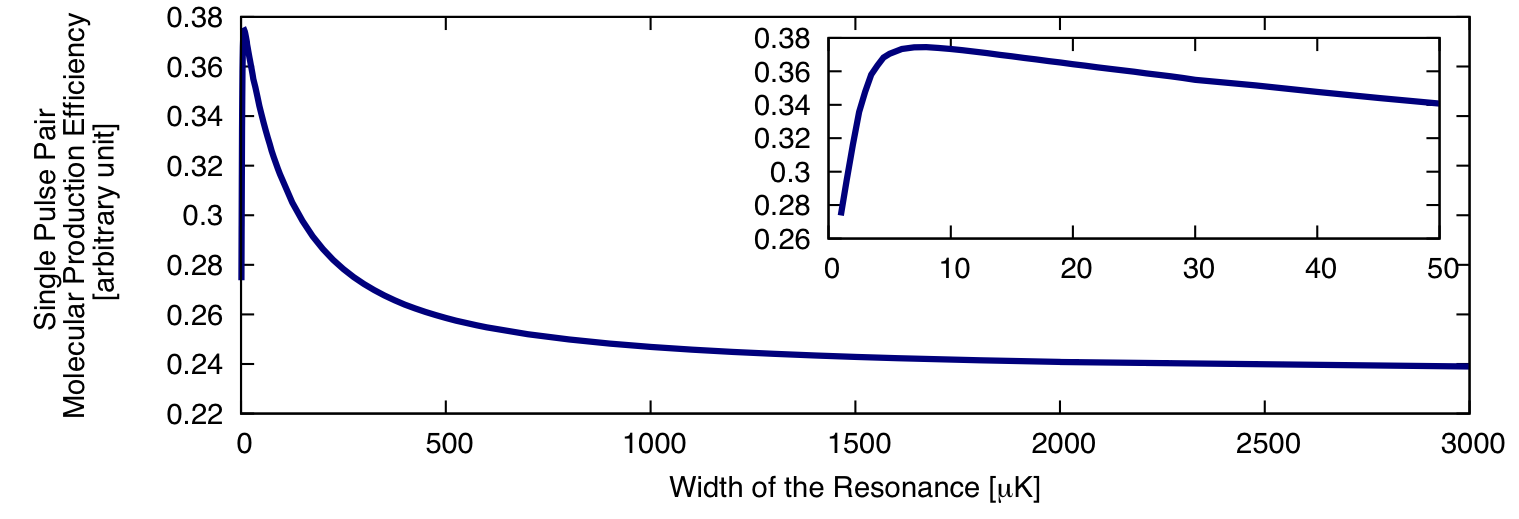}
\caption{The time-averaged molecular production yield for a single pulse pair for an atom ensemble at $100\mu$K, calculated by integrating the delay
plots of Fig. \ref{delay}.}
\label{improv}
\end{figure}

In order to obtain the delay times averaged molecular production yield for an
atomic ensemble we need to calculate the area under the transfer-yield
curves of Fig. \ref{delay}.
Figure \ref{improv} displays the dependence of the delay times
averaged yield for various resonance widths.
We first note that the yield changes relatively slowly for
resonance width larger than $1000\mu$K. This is because in this case the
resonance width by far exceeds $\delta_0,$ the energetic
spread of the initial atomic ensemble and we approach the
flat-continuum limit. As the width of the resonance drops to a few $\mu$K,
the molecular production yield rises to a maximum value, but drops
significantly due to spontaneous emission
for yet narrower resonances. Thus there exists an
optimal resonance width for which the molecular production yield is
maximal. Comparing the optimal molecular production yield, obtained for
a (narrow) resonance value of $\sim 8\mu$K, with the yield in the
flat-continuum limit, we see an improvement factor of 1.56.

\subsection{Scaling behaviour with ensemble temperature}
We now explore, as was done in Ref. \cite{vardi}, how the process
varies as the average ensemble energy, $E_0,$
and energy spread, $\delta_0,$ are scaled down by a factor of $s>1$,
i.e., $E_0\rightarrow {E_0\over s}$, and
$\delta_0\rightarrow {\delta_0\over s}.$ In ref. \cite{vardi},
we showed that the equations are invariant to
this scaling provided the peak time was scaled up by the same factor
$t_0\rightarrow st_0,$ and
the initial wave packet amplitude is scaled as
$b_E(0;E,t_0) \rightarrow \sqrt{s} b_E\left(0;{E\over s},st_0\right)$.
We now consider the effect, in addition to the above, of
scaling the resonance shape as
$\mu_{2,E}(E,\Gamma_{res})\rightarrow\mu_{2,E}\left({E\over s},{\Gamma_{res}\over s}\right)$. In order to match the spectral profile to the scaled $b_E(0)$,
we need to scale up the centre frequencies and durations of the
two pulses by the same $s$ factor.
Since we can choose the intensity (amplitude) of the fields,
we scale $\epsilon_1(t)\rightarrow {\epsilon_1(ts)\over s}$ and
$\epsilon_2(t)\rightarrow {\epsilon_2(ts)\over\sqrt{s}}$ \cite{vardi}
As a result, the source function scales like
$f_{source}(t,t_0, \Gamma_{res}) \rightarrow
s^{-1}\cdot f_{source}(st, st_0, {\Gamma_{res}\over s})$.
The above scaling leaves the dynamical
equations (Eqs. (\ref{b1dyneq}-\ref{B1dyneq})), essentially unchanged,
except for the spontaneous decay rate which cannot be scaled.
As we scale the relevant times by a factor of $s$, the deleterious
effect of the spontaneous emission becomes more and more pronounced.

\begin{figure}[top]
\centering
\includegraphics[width=1\columnwidth]{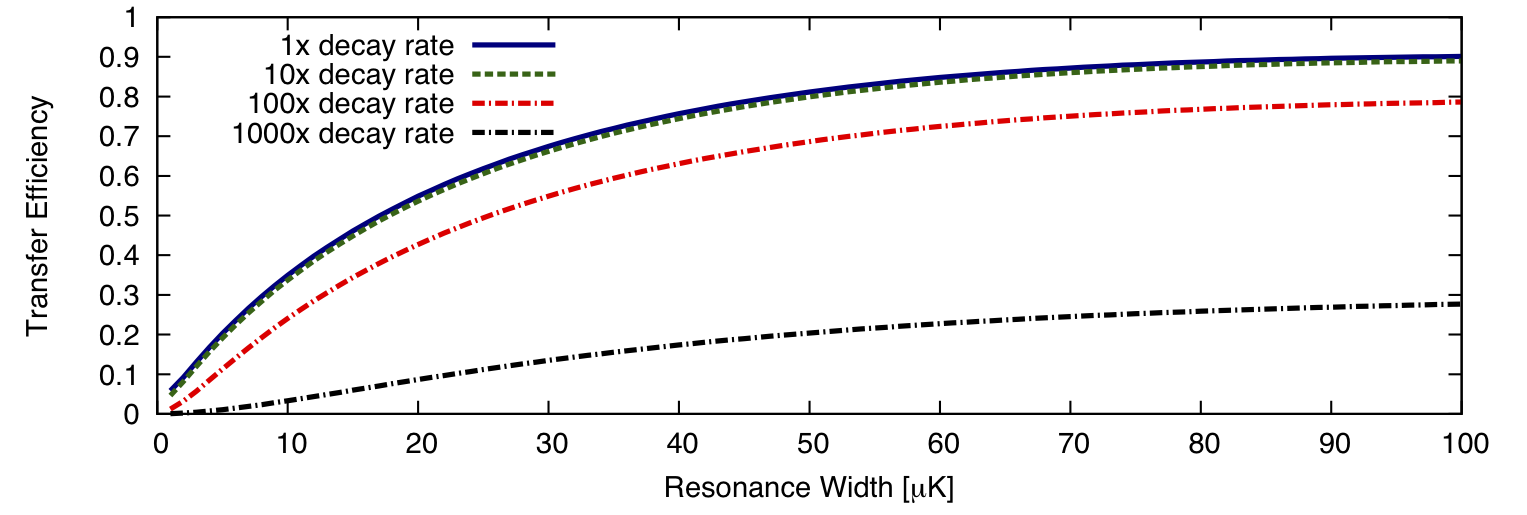}
\includegraphics[width=1\columnwidth]{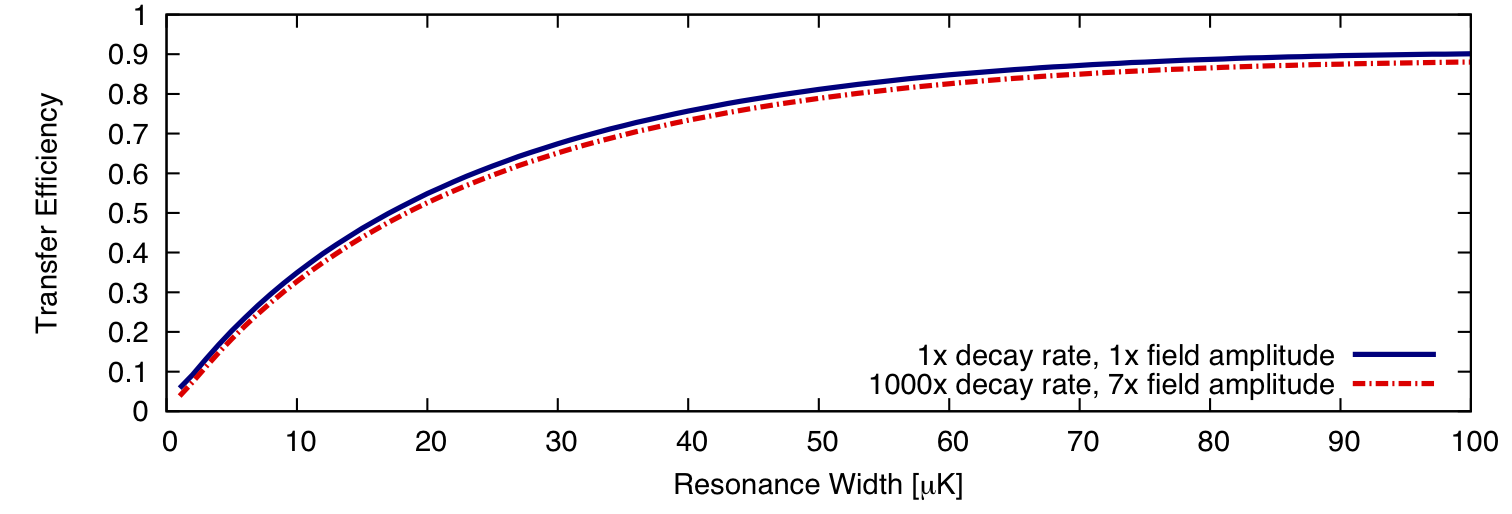}
\caption{{\bf Top panel:} The single collision photoassociation yield
$|b_1(t\rightarrow\infty)|^2$ as a function of resonance width, for four different values of the spontaneous decay rate. {\bf Bottom panel:} The same plot at
stronger laser intensities. The photoassociation yield now
becomes insensitive to the spontaneous decay rates.}
\label{freedecay}
\end{figure}

In Fig. \ref{freedecay} top panel we display the dependence of the single collision
photoassociation yield on the spontaneous decay rate. Note that a change in one order of magnitude for the spontaneous decay rate only affects our results negligibly.
When we compare the results to those
displayed in Fig. \ref{efficiency} top panel,
where the transfer yield is plotted as a function of the resonance width,
we see that the transfer yield  is not greatly affected
at ensemble temperatures of a few $\mu$K to a few $10^2\mu$K.
As the ensemble temperature goes down by three orders of magnitude,
the single collision transfer efficiency goes down too, by $\sim 67\%$ of its
original value. The effect is more pronounced for narrower resonances,
because the longer interaction times enhance the effect of the spontaneous
decay. However, as shown in the lower panel of Fig. \ref{freedecay},
it is possible to combat the effect of spontaneous decay at very low
temperatures, e.g. in nK range, by increasing the amplitude (intensity)
of both laser fields.

One can summarize these results by saying that the optimal resonance width
is always $\sim 8\%$ of the
ensemble temperature, and that the optimal molecular production efficiency
by ARPA is $\sim 56\%$ higher than that of the wide resonance
(flat continuum) case.

\subsection{Thermalization}
At long times the cumulative action of many pulses can change a region
(or regions) in phase space corresponding to the recombining atoms, thereby
affecting the initial wave packet amplitude $b_E(0).$
However, an atomic ensemble can thermalize sufficiently fast, on the order
of milliseconds, to yield the typical atomic trap
setting of $100\mu$K temperature and $10^{11}$/cm$^3$ density \cite{kirk}.
This means (depending on the repetition rate of the pulses), that after a
few thousand $\mu$s pulse-pairs, the atomic ensemble can thermalize back to
its original phase-space distribution, re-validating the ensemble-averaged form
we used for $b_E(0)$.

According to previous estimates \cite{vardi,coldbook}, the total number
of pulse pairs needed to transfer an entire atomic ensemble of density
$10^{11}/$cm$^3$ is around of $10^7$. Therefore a few thousand pulses is
indeed a very small fraction of total number of pulses needed,
and the thermalization is fast comparing to the ensemble size molecular
conversion time.

One is also inclined to pose the practical question of how to hide the newly formed molecule in state $|1\rangle$ from subsequent pulse pairs. In accordance with more detailed discussions in Ref \cite{vardi, coldbook}, this can be done, for example, by allowing the newly formed molecules in state $|1\rangle$ to ``leak" away from the laser focus, which is possible because they react differently from the atoms to the confining laser frequency. A molecular trap can then be placed just below the atomic trap.

\section{The ARPA yield dependence on the phase-space density and
comparison to magneto-association.}

Following Refs. \cite{vardi,zhenia07}, we now present
a detailed calculation of ARPA efficiency in a thermal ensemble.
In order to estimate the fraction of atoms photoassociated per
pulse-pair, we multiply $P(E)$, the single collision photoassociation
probability at energy $E$, by the number of
collisions experienced by a given atom while the pulses are on. This is
equivalent to averaging over all possible values of $t_0$ as performed above.

The number of collisions during the pulses is calculated as follows:
at a given energy $E$, the velocity of a given atom is
$v=(2E/m)^{1\over2}$ and the distance traversed by it during a
pulse of $\tau_{laser}$ duration is $v\tau_{laser}$. The
cross-section for collision is $\pi b^2$ where $b$ is the impact
parameter. For $s$-wave collisions, the semiclassical estimate is
$b=\hbar /2 p=\hbar/(2\sqrt{2mE})$. Hence, the number of
collisions experienced by the atom during the two pulses is $n=N
\pi b^2 v\tau_{laser} /V$ where $N$ is the number of atoms in the
trap, and $V$ is its volume. Putting all this together we obtain
that the fraction of atoms photoassociated per pulse-pair is
\begin{equation}\label{estimate}
f(E)= {P(E)\pi N\tau_{laser}\over 4 V m^{3/2}(2E)^{1\over 2}}~.
\end{equation}
Estimating the photoassociation yield for the case of flat
continuum, we can set $P(E)\simeq 1$ \cite{zhenia07}, and assume
that all collisions occur at the temperature of the relative
motion $T_{rel}=2 T$ \cite{TemperatureComment}. Further, when
optimizing the yield for an atomic ensemble, we must choose
$\tau_{laser} \simeq 2\pi\hbar / kT_{rel}$ because the bandwidths of the
pump and dump pulses should match the energy
spread in the ensemble. Thus we obtain
\begin{equation}\label{PAyield}
f(T)\simeq {\pi^2 N \hbar^3 \over  V (2mkT_{rel})^{3/2}}~.
\end{equation}
As pointed out above, a narrow resonance can enhance this fraction
by a factor of $0.56$, i.e., for a narrow resonance $f^{narrow}(T)=1.56f(T),$
with $f(T)$ given by Eq. (\ref{PAyield}).

We now consider the yield of magneto-association. In this process
a time-varying external magnetic field is ``swept" in magnitude, thereby
moving $E_s,$ the position of the Feshbach resonance of interest,
to lie below $E_{th},$ the onset of the continuum. In this way the
Feshbach resonance is stabilized to become a ``Feshbach
molecule'' \cite{magneto1, magneto2, magneto3, magneto4, magneto5}.
The magneto-association is then followed by a
traditional 3-bound states STIRAP \cite{magneto4,magneto5}.

The efficiency of this scheme is limited by the yield of the first step.
In this step two atoms may
form a molecule if prior to sweeping the magnetic field they are
within, approximately, $\Theta_{association}=\left(\hbar/2\right)^3$
volume of phase space from each other \cite{Wieman-Fesh-PRL05}.
Therefore at low to moderate phase space
densities, the probability for a given atom to participate in a
magneto-association process is
\begin{equation}
\label{MAyield} f(T)\simeq N\,\Theta_{association}/ \Theta_{whole}
= \frac{N\,\hbar^3}{8 V (2 m kT_{rel})^{3/2}}
\end{equation}
where $\Theta_{whole} = V\times(2 m k T_{rel})^{3/2}$ is the
single-particle phase space volume at temperature $T$, and
$V$ is the trap volume.

Comparing Eqs. (\ref{PAyield}) and (\ref{MAyield}) we see that
the ARPA yield scales with temperature in exactly the same
fashion as the magneto-association yield. However, in absolute numbers
our estimates are
that at low to moderate phase space densities the ARPA yield is
$1.56\times 8\pi^2 \simeq 120$ times higher than the
magneto-association yield. These findings strongly suggest that an
experimental investigation of ARPA at sub-$\mu$K temperatures
is warranted.

\section{Conclusions}
In this paper we have shown that
Adiabatic Raman Photoassociation of ultracold atoms proceeding via collisional
resonances is an efficient way of producing
ultracold diatomic molecules in deeply bound states.
We have done that by replacing the
resonance-dominated molecular continuum by a discrete set of
``effective modes'' acting like a set of resonances.
Though when the scattering resonance width is narrow it
covers a smaller region in phase space (relative to the case of
wide resonances), resulting in a drop of
the single collision transfer efficiency, this drop is amply compensated
for by the (as much as an order of magnitude) longer durations
at which the photoassociating pulses can effectively act. In this way
each pair of (pump and dump) laser pulses
can act on more colliding atoms. The overall effect is
that the narrow-resonances molecular production yield
can be as much as $\sim 56\%$ higher
than the wide resonances yield. For atomic temperatures in the $\mu$K range,
we find that the optimal conditions are attained
for resonances whose widths are about 8 \% of the ensemble temperature.
We have also shown that the efficiency of the ARPA scheme compares
favourably with the efficiency of magneto-association, with
the yields of both schemes scaling with temperature in exactly the same manner.
We have demonstrated that the ARPA process is a projective quantum
measurement by the pulses of the initial continuum wave packet. This feature
is a result of the single collision
transfer efficiency being proportional to the
degree of overlap between a function set by the pulses and the initial
wave packet.

Future applications will deal with time-dependent resonances.
We envision combining ARPA with a dynamical sweep of
the Feshbach resonance across the threshold energy range.
As the sweep will render the resonances narrower, the
laser pulses will be made narrower so as to transfer
the atomic gas into molecules in an optimal piecewise manner.

\vskip .3truein

\noindent
{\bf Appendix: ARPA as a projective measurement of the initial continuum wave
function.}

We showed in reference \cite{coldbook} that
if the continuum is flat then ARPA implements a projective measurement
of the initial wave function of two colliding atoms. Basically, the profiles
of the laser pulses $\epsilon_n(t)$ define a wave form
$f^{(ARPA)}$ that is adiabatically coupled to the target state $\1$. An
initial scattering state which overlaps
$f^{(ARPA)}(t)$ well will undergo population transfer to $\1$, while a state
orthogonal to $f^{(ARPA)}$ will not. By controlling the laser pulse
profiles and implementing ARPA one is essentially measuring the
wave function of the colliding atoms.

In this Appendix we extend the treatment to a resonance-dominated continuum,
and relate the effect of collision resonances with our ability to
control $f^{(ARPA)}(t)$. We start by noting that in the adiabatic
limit the solution of the equations of motion
(\ref{SVCA-bdot},{\ref{SVCA-H}}) is of the form \cite{vardi,coldbook}
\begin{eqnarray}
b_1(t) &=& i\, \c(t) \int_0^t dt' \exp\left[i\, \int_{t'}^t
          \E_+(t'')\,dt''\right] \s(t')~
{f_{source}(t')}\nonumber \\
        &-&
        i\, \s(t) \int_0^t dt' \exp\left[i\, \int_{t'}^t
          \E_-(t'')\,dt''\right] \c(t')~
{f_{source}(t')}
          \label{b1oft} \\
b_2(t) &=& i \,\s(t) \int_0^t dt' \exp\left[i\, \int_{t'}^t
          \E_+(t'')\,dt''\right] \s(t')~
{f_{source}(t')}  \nonumber \\
        &+&
        i\, \c(t) \int_0^t dt' \exp\left[i\, \int_{t'}^t
          \E_-(t'')\,dt''\right] \c(t')~
{f_{source}(t')}~.
          \label{b2oft}
\end{eqnarray}
where
\begin{equation}
\E_\pm(t) = \frac{1}{2} \left\{i\G_{eff}(t) \pm
\sqrt{4|\W_{12}(t)|^2 - \G_{eff}(t)^2} \right\} \ ,
\label{Epm}%
\end{equation}
and
\begin{equation}
\tan\theta(t) = \E_+(t)/\W_{12}(t) \ .
\label{cs-def}
\end{equation}

The final yield of the ARPA process is defined as the probability
$P_1=|b_1(t\rightarrow\infty)|^2$. Using Eqs. (\ref{Epm}) and
(\ref{cs-def}), we see that $\c(t\rightarrow\infty) = 0,$ and the
excited bound state amplitude $b_2(t\rightarrow\infty)$ indeed vanishes.
Substituting
$\c(t\rightarrow\infty) = 0$ and $\s(t\rightarrow\infty) = 1$ in
Eq.~(\ref{b1oft}) we obtain that
\begin{equation}%
   b_1(t\rightarrow\infty) =      %
   \int_0^{\infty}  f^{(0)}_{\rm ARPA}(t) f_{source}(t) \,dt          %
  \equiv \langle f^{(0)}_{\rm ARPA}|f_{source}\rangle_t
\label{ProjectionPlus}\end{equation}
where
\begin{equation}\label{Farpa0}
{f}^{(0)}_{\rm ARPA}(t) = -i  \exp\left[i\, \int_{0}^{t}
          \E_-(t')\,dt'\right] \c(t).
\end{equation}

Thus the photoassociation amplitude $b_1(t\rightarrow\infty)$ is
given as the projection of the source function $f_{source}$ onto
the specific wave form ${f}^{(0)}_{\rm ARPA}$ whose shape
is controlled by the
amplitudes and the phases of $\W_D(t)$ and $\W_P(t)$. Wave packets
that are orthogonal to ${f}^{(0)}_{\rm ARPA}$ do not photoassociate in
the ARPA process, while the ones that project well onto
${f}^{(0)}_{\rm ARPA}$ do. By tailoring the amplitudes and phases of
the laser pulses, one can choose which continuum waveform is
transferred into the target state \cite{coldbook}.

In the main part of the paper we have shown
that resonance dominated ARPA is most efficient
when the resonance is narrow and the pump pulse has a narrow bandwidth
relative to the initial ensemble temperature.
These arguments allow us to replace in
(\ref{fsource}) ${E_{th}}$ by $-\infty$ and obtain, via the
convolution theorem,
\begin{equation}
\label{fW}
f_{source}(t)  = \epsilon_2(t)(f_0\times W) (t)\equiv %
\epsilon_2(t) \, \int_{-\infty}^\infty W(\tau) f_0(t-\tau)\,  d\tau~,
\end{equation}
where
\begin{equation}\label{W}
W(t)  = {1\over 2\pi}\int_{-\infty}^\infty  \mu_{2,E}(E)
\,e^{i\Delta_E t} \,dE,
\end{equation}
and
\begin{equation}\label{f0} f_0(t)  = \int_{-\infty}^\infty  b_E(0)
\,e^{i\Delta_E t}\,dE.
\end{equation}
$f_0(t)$ is the phase-space {\it envelope} \cite{zh-envelope} of the
initial wave packet of continuum states.
Semiclassically it corresponds to the incoming wave function
as a function of time $t,$
measured at the turning point of a classical trajectory of energy $E_0$.
Positive values of $\tau$ - the time variable in the convolution integral of
Eq. (\ref{fW}) - correspond to an outgoing motion, and negative
values -  to an incoming motion \cite{zh-envelope,coldsemiclassical-comment}.

The function $W(t),$ the ``FC window'', describes the residence time of
the system in the ``FC region,'' the spatial region for which
the FC factors are substantial. The temporal
width $\D\tau_W$ at which $W(\tau)$ is
substantial, corresponds to $\D R,$ the spatial extension
of the FC window.
In the flat continuum case, $W(\tau)=\mu\delta(\tau)$, and the time of
residence in the FC region is zero.
In contrast, a narrow resonance can cause the system to be
greatly delayed in the FC window.
In that case, an incoming continuum wave packet does not leave the
Franck-Condon region right after entering it, but rather dwells
there for the time given by the width $\D\tau_W$.

We now consider the role of the continuum structure. Combining
Eqs.(\ref{fW}) and (\ref{W}) with (\ref{ProjectionPlus}), and
introducing $t'=t-\tau$ we obtain
\begin{equation}
\label{proj-intob1-2}%
   b_1(t\rightarrow\infty) =      %
   \int_{-\tau}^{\infty}  {f}^{(W)}_{\rm ARPA}(t') {f^{(0)}}(t') \,dt'
   \equiv \langle {f}^{(W)}_{\rm ARPA}|{f^{(0)}}\rangle_t         %
   ~,
\end{equation}
where
\begin{equation}\label{FarpaW}
{f}^{(W)}_{\rm ARPA}(t') = \epsilon_2(t') \int_t^\infty
{f}^{(0)}_{\rm ARPA} (t'+\tau)
 W(\tau) \,d\tau.
\end{equation}
Thus the shape of the wave form scooped from continuum by a laser
pulse pair is equally defined by the laser pulses and by the
continuum structure encoded in the window function $W(\tau)$.

Expanding the bound continuum transition spectrum into effective
modes (Eq.(\ref{resonances})) we have
\begin{equation}
W(\tau) = - \sum_s {\mu_s \G_s \over 2}\, \exp\left[\left(i (E_2-\w_2-E_s) -{\G_s\over2}  \right)\tau\right]\,\theta(\tau)  %
\label{W-lorentzian}
\end{equation}
where $\theta(\tau)$ is the Heaviside function. Therefore
\begin{equation}\label{ProjectionPlusres}
{f}^{(W)}_{\rm ARPA}(t) = -\epsilon_2(t) \sum_s  {\mu_s \G_s \over
2} \int_0^\infty d\tau\, {f}^{(0)}_{\rm ARPA} (t+\tau) e^{\left(i
(E_2-\w_2-E_s) -{\G_s\over2} \right)\tau}.
\end{equation}
Equations (\ref{ProjectionPlus},\ref{ProjectionPlusres}) present
the main result of this section. They show that, similar to the
case of a flat continuum, the profiles of the pump and dump pulses
define the shape of coherent wave forms which can be transferred
from the continuum into the target state. However the dwelling of the
wave function due to the resonances decreases the ability to control these
wave forms.
If due to the resonances the dwell
time $\D\tau_W$ exceeds the durations of the laser pulses, then by
Eq.(\ref{FarpaW}) we know that in addition to photoassociating atoms
that arrive at the FC at $t_0,$ there is a
non-negligible probability to photoassociate atoms which get there
before or after $t_0.$


\section{Acknowledgements}

Support by NSERC Discovery Grant, by a Major Thematic Grant from UBC's
Peter Wall Institute for Advanced Studies,
and by the US DoD DTRA program are gratefully acknowledged. E.S. acknowledges the Institute of Theoretical Atomic, Molecular, and Optical Physics (ITAMP) for support during a visit to ITAMP facilities.


\begin{thebibliography}{1}
\bibitem{coldsemiclassical-comment}
At ultracold temperatures the semiclassical point of view
underlying the above carrier-envelope interpretation of the wave
function is invalid for long range van-der-Waals interactions, i.e. at
interatomic distances from several up to hundreds of atomic units.
However, our approach predicts what happens to the wave function
after the photoassociation is complete, at large interatomic
distances where the semiclassical expression for the wave function is correct.%
\end{thebibliography}
\end{document}